\documentclass[aps,show pacs,singlecolumn,preprint,amsmath,amssymb,superscript address,prx]{revtex4-1} 

\usepackage{graphicx}
\usepackage{mathtools}
\usepackage{color} 

\begin{document} 
\title{Scalable, ultra-resistant structural colors based on network metamaterials}

\author{Henning Galinski$^{1,2\ddag}$, Gael Favraud$^{3,\ddag}$, Hao Dong$^{1,2}$, Juan S. Totero Gongora$^3$, Gr\'egory Favaro$^4$, Max D\"obeli$^5$, Ralph Spolenak$^2$, Andrea Fratalocchi$^{3,\dag}$ and Federico Capasso}
\email{capasso@seas.harvard.edu, $^\dag$ andrea.fratalocchi@kaust.edu.sa} 
\thanks{$^\ddag$ These authors contributed equally to the work.}
\affiliation{Harvard School of Engineering and Applied Sciences,Harvard University, 29 Oxford Street ,Cambridge, USA\\
$^2$Laboratory for Nanometallurgy, ETH Zurich, Vladimir-Prelog-Weg 1-5/10, Zurich, Switzerland\\
$^3$PRIMALIGHT, King Abdullah University of Science and Technology (KAUST), Thuwal 23955-6900, Saudi Arabia\\
$^4$CSM Instruments SA, CH-2034 Peseux, Switzerland\\
$^5$Ion Beam Physics, ETH Zurich, Zurich, Switzerland}

\date{\today} 
\begin{abstract} 
Structural colours have drawn wide attention for their potential as a future printing technology for various applications, ranging from biomimetic tissues to adaptive camouflage materials. However, an efficient approach to realise robust colours with a scalable fabrication technique is still lacking, hampering the realisation of practical applications with this platform. Here we develop a new approach based on large scale network metamaterials, which combine dealloyed subwavelength structures at the nanoscale with loss-less, ultra-thin dielectrics coatings. By using theory and experiments, we show how sub-wavelength dielectric coatings control a mechanism of resonant light coupling with epsilon-near-zero (ENZ) regions generated in the metallic network, manifesting the formation of highly saturated structural colours that cover a wide portion of the spectrum. Ellipsometry measurements report the efficient observation of these colours even at angles of $70$ degrees. The network-like architecture of these nanomaterials allows for high mechanical resistance, which is quantified in a series of nano-scratch tests. With such remarkable properties, these metastructures represent a robust design technology for real-world, large scale commercial applications. 
\end{abstract} 

\maketitle 


\section*{Introduction}
Billions of years ago, the green algae originated life, changing the face of the earth from grey to green and paving the way for the life-forms we see today \cite{Yoon1}. Ever since, living organisms have extensively used colours for a variety of purposes, ranging from communication to self defense, from reproduction to camouflage \cite{Bradbury1}.
The enormous variety of colors, such as the sapphire blue wings of the Morpho butterfly~\cite{potyrailo2007morpho,Kinoshita2008} or the thermochromic coloration of the chameleon~\cite{Teyssier2015}, has stimulated the interest of researchers dating back to the 17$^{\text{th}}$ century, when Hooke reasoned about the origin of colour in brilliant feathers of peacocks and ducks~\cite{Hooke1665}. A large part of these colours do not originate from pigments or dyes but is `structural', resulting from the interaction of light with self-assembled structures of living organisms~\cite{Kinoshita2008,huang16:_harnes_struc_darkn_in_visib,vukusic07:_brill_whiten_in_ultrat_beetl_scales}.\\
\indent The engineering of structural colors from artificial photonic structures attracted a conspicuous interest in research, due to the many applications that can potentially be opened by this technology \cite{SeagoS165,Kats3,Kats2,Schlich1,Zhang2015,Joannopoulos2015,Kumar2012,C5NR00578G,Zi28102003,Teyssier2015}. Structural colours based on photonic crystals and metamaterials have been widely explored, showing very promising results such as the possibility to create colours at the diffraction limit \cite{Kumar2012}. A major challenge is to overcome problems of limited scalability and lack of robustness, which affect the technology of photonic crystals and classical metamaterials. It is therefore highly desirable to investigate new approaches that can transform these initial breakthroughs into real world applications.\\
\indent In the following we describe a new biomimetic material that overcomes the aforementioned challenges, introducing a new type of structural coloration that is highly scalable and extremely robust. This nanomaterial takes inspiration from the structural coloration of the \textit{Cotinga maynana}, a South American bird~\cite{Dyck1}. The non-iridescent blue colour of its feathers is produced by an aperiodic nanoporous keratin network with a typical feature size smaller than $200$~nm. This lightweight network has extraordinary optical properties, which cannot be explained by classical scattering and still lacks a convincing physical explanation~\cite{Prum1998}. The interaction of light waves with complex materials already reported a series of fascinating dynamics, ranging from energy harvesting to ultra-dark nanomaterials and beyond \cite{liu13:_enhan_energ_storag_in_chaot_optic_reson,coluccio_detection_2015,Kats2,Cao1,Antosiewicz1,Verre1,PhysRevLett.79.4562,shalaev13:_nonlin_optic_of_random_media,liu_triggering_2015,huang16:_harnes_struc_darkn_in_visib}.
By using the \textit{Cotinga maynana} feathers as a design principle, we create complex nano-photonic structures that combine a cellular metallic network~\cite{Bauer2016,Hodge2006} with subwavelength coatings made by loss-less dielectrics. This material combination provides significant advantages for real world applications: it is suited for large-scale fabrication, light-weight and mechanically robust, combining the high yield strength to low density ratio of a cellular metallic network with the resistance to wear that alumina offers~\cite{Mayer2003}. Optically, the interface of such a metallic nanoscale network and the lossless dielectric can be considered as electromagnetically ``weakly" rough and an inhomogeneous mixture of dielectric/metal and dielectric/air regions. In this scenario, the component of the wavevector parallel to the interface is not conserved, resulting in a highly spatially dependent electromagnetic response. Taking advantage from such a complex light-matter interaction, we here illustrate how to create colours with remarkable properties.

\section*{Materials and Methods} 
\subsection*{Sample Preparation and Characterization}
Pt-Y-Al layers of 300 nm thickness were deposited at room temperature by magnetron co-sputtering
onto SiNx/Si substrates that were pre-cleaned using isopropanol and acetone. Subsequently, the films have been dealloyed in 4M NaOH at room temperature for 60 s and rinsed in deionised water afterwards. The morphological analysis of the samples was studied via scanning electron microscopy (SEM) assisted by focused ion beam etching (FIB). The compositional analysis was performed by Rutherford backscattering spectrometry. Detailed information is given in the supplementary information.
In this work the Savannah atomic layer deposition (ALD) from Cambridge NanoTech was used to deposit Al$_2$O$_3$ coatings on the dealloyed metal nanowire networks. During the ALD deposition of Al$_2$O$_3$ a pulse time of $0.15$~s and a purge time of $30$~s
for both TMA and water was used. The base pressure was $500$~mtorr and the working temperature was $250^{\circ}$~C. The growth rate was around $0.1$~nm  per cycle.
For the colored graphic art illustrated in Fig. \ref{art}, the $60$~nm thick Al$_2$O$_3$ film has been deposited via radio frequency (RF) sputtering 
at room temperature using an ATC sputtering tool (AJA International).
The electromagnetic reflectance of the coated samples has been measured using a variable-angle spectroscopic ellipsometer from J.A. Woollam Co and a NanoCalc thin film reflectrometry setup (Ocean Optics Inc.). The dielectric constant of the Al$_2$O$_3$ coating deposited by ALD has been determined using a Cauchy model by analyzing a $53$nm thick Al$_2$O$_3$ coating deposited on a Si wafer. The scratch tests have been performed using an Anton Paar TriTec Nano Scratch Tester (NST).

\section*{Results and Discussion}
\subsection*{Material design and color characterization}
We have selected dealloying to assemble a nanoscale metallic network with controllable features. This method, first proposed by Raney to synthesise metal catalysts \cite{Raney1}, utilizes the selective dissolution of the less noble constituent of an alloy during wet-etching. In our experiments, 300 nm thick Pt$_{.14}$Y$_{.06}$Al$_{.80}$ thin films are deposited on an amorphous Si$_3$N$_4$/Si substrate. While immersing the film in a 4 molar aqueous solution of NaOH for $60$s, the less noble Al in the Pt-alloy thin film is subsequently removed and the remaining metal reorganises to a network with an open porosity, mimicking neural architectures. Characteristic geometrical features of the network can be altered by changing the etching time, the etchant concentration or the initial composition of the thin film~\cite{Galinski2,Supansomboon2014,AbdelAziz2015,Chen2013,Qi2015}.\\
In a second step the nanomaterial is coated with an ultra-thin layer of Al$_2$O$_3$ using atomic layer deposition (ALD). The coating thickness is increased stepwise in a range from $7$ to $53$~nm. We characterise the growth of the subwavelength Al$_2$O$_3$ coatings by Rutherford backscattering spectroscopy and focused ion beam assisted scanning electron microscopy (see supplementary information). A three-dimensional image of the Pt$_{.56}$Y$_{.26}$Al$_{.18}$ network, experimentally obtained using Focused Ion Beam (FIB) thin-film tomography, is displayed in Fig. \ref{pal}a.\\
\indent In a first series of experiments, we characterise the optical response of the network metamaterial for different thicknesses of the dielectric layer Al$_2$O$_3$. These experiments unveil a very interesting mechanism of structural color generation from the nanowire network, as shown in Fig. \ref{pal}b. By changing the coating thickness, we observe the generation of a multitude of colours spanning from yellow, orange, red, to finally blue. When the same coatings are deposited on a dense PtYAl metal thin film, conversely, no particular colour is produced (see Supplementary information and Supplementary Fig. 5). The colours generated in the metallic network are ultra-saturated and go even slightly beyond the Red Green Blue (RGB) gamut in CIE chromaticity diagram (Fig.~\ref{pal}c).\\
\indent In order to illustrate that these colours are consistently generated by varying $\mathrm{Al_2O_3}$ layer thickness, we compare experimental results with theoretical predictions based on Finite-Difference Time-Domain (FDTD) simulations. For the latter, we use a 2D section of the FIB tomography of the sample illustrated in Fig. \ref{pal}a.
Our FDTD simulations, shown in Fig. \ref{pal}c as a dotted line, well reproduce the experimental results, demonstrating indeed the possibility to achieve such large variety of colours by tuning the thickness of the $\mathrm{Al_2O_3}$ layer. Figure \ref{pal}b reports experimental images of samples characterised by different thicknesses of $\mathrm{Al_2O_3}$. Quite remarkably, despite of the metallic nanoscale network that exists below the $\mathrm{Al_2O_3}$ layer, the samples show a very uniform colour in all different configurations. A comparison with FDTD calculations is provided in Fig. \ref{pal}d, which illustrates the color palette that can be generated by the system when the thickness of $\mathrm{Al_2O_3}$ increases.\\
 To emphasise that the generation of structural colours in these nanoplasmonic structures can be achieved by various deposition techniques, we also fabricated a structural coloured graphic arts by using physical vapor deposition. Figure \ref{art} depicts an example created by combining a dealloyed network metamaterial with an RF-sputtered $60$~nm thick $\mathrm{Al_2O_3}$ coating and simple photo-lithography using a Heidelberg $\mu$PG$501$ optical direct writing system. The bicolored graphic art combines highly uniform structural colour (blue) with a metallic white colour (dense film).

\subsection*{Robustness of structural colours from metamaterial networks}
To quantify the mechanical robustness of these colors, we resort to nano-scratch resistance testing (Fig. \ref{wear}a), which is an ideal technique to characterise the adhesion failure of coatings. A detailed description of the experimental procedure we used is given in the supplemental material. Figure \ref{wear}b reports optical micrographs of four representative nano-scratch tests. The wear resistance of a dense PtYAl film with and without a 28 nm thick Al$_2$O$_3$ coating is compared to a porous nanoscale Pt network coated with 28 nm and 53 nm of Al$_2$O$_3$, respectively. The critical load causing delamination of the coated network metamaterial is almost 2 times higher than the corresponding dense metallic film (Fig. 2b) and $20\%$ higher than the dense metallic film coated with 28 nm thick Al$_2$O$_3$. Considering the $53\%$ of porosity in the nanoscale network, the observed increase in wear resistance is remarkable and indicates an enhanced strength-to-density ratio~\cite{Ashby2006} that goes alongside with a significant reduction of overall weight of the coating.\\
\indent Figure \ref{spec} illustrates s-polarised reflectivity spectra at normal (Fig. \ref{spec}a) and oblique (Fig. \ref{spec}b-e) incidence for different alumina coating thickness. Figure \ref{spec}a shows that the generation of colours originate from a large red shift of the reflectivity response of the nanomaterial, observed when the $\mathrm{Al_2O_3}$ layer changes thickness. The corresponding FDTD results are reported in Fig. \ref{spec}b. FDTD simulations quantitatively well reproduce experimental results, confirming the principal role of the $\mathrm{Al_2O_3}$ coating layer in red-shifting the spectral response of the material. A small variation of only $30$ nm in the $\mathrm{Al_2O_3}$ thickness shifts the reflectivity minimum of approximatively $350$ nm. Reflectivity spectra of the material are stable and do not show significant variations up to incident angles of $70^\circ$, which still provide reflectivity minima as low as $<1$\% (Fig. \ref{spec}b-e). The mean angular dispersion of the reflectance minimum has been determined from the reflectance spectra obtained by ellipsometry. The mean angular dispersion is independent of the coating thickness and the reflectance minimum blue shifts with $-1.0\pm 0.3$~nm/$^\circ$. These experiments show that the structural colours observed in Fig. \ref{pal} are non-iridescent, i.e robust against large changes of the incident angle. 

\subsection*{Structural colour generation from localised surface states in complex epsilon-near-zero materials} 
In this section we analyse in more detail the mechanisms by which structural colours are created and observed in the metallic network of Fig. \ref{pal}. When polychromatic light impinges on the structure of Fig. \ref{pal}a, the interaction between light and matter generates surface plasmon polaritons (SPP) \cite{maier2010plasmonics}, which are surface waves localised at the metal-dielectric interface of the structure \cite{huang16:_harnes_struc_darkn_in_visib}. In our samples (Fig. \ref{plas}a), due to a strongly disordered metallic profile, the motion of SPP develops along complex trajectories in space (Fig. \ref{plas}a, inset). It is convenient to study this motion in a new curvilinear system, whose axes are parallel to the spatial trajectories of SPP. To this extent, we introduce a new set of coordinates $(\psi,\phi)$, which are conformal to the disordered surface of the metal. Figure \ref{plas}a shows how these coordinates appear in the original space $(x,y)$, while Fig. \ref{plas}b shows how the original structure appears in the space $(\psi,\phi)$, which we identify as the \emph{plasmonic reference}. In the plasmonic reference, the motion of surface plasmons is extremely simple, and composed of straight lines at $\psi=0$ (Fig. \ref{plas}b, inset). When we change spatial coordinates in any electromagnetic system, Maxwell equations stay invariant if we introduce an inhomogenous refractive index distribution that makes the two reference systems equivalent \cite{Pendry03082012,leonhardt2006optical}. The pseudocolor plot in Fig. \ref{plas}b shows the spatial distribution of the inhomogenous index $n(\psi,\phi)$, computed by using transformation optics \cite{Pendry03082012,leonhardt2006optical}. The index $n(\psi,\phi)$ is associated to the coordinate transformation introduced in Fig. \ref{plas}b and acts as a counterpart of the metallic geometry of Fig. \ref{plas}a, which does not exist in Fig. \ref{plas}b as the metal surface is flattened out. The two structures of Fig. \ref{plas}a and Fig. \ref{plas}b are exactly equivalent: when light propagates in one or another, it follows the same dynamics. This is an exact result of Maxwell equations that contains no approximation. This also implies that when light impinges on the structure of Fig. \ref{plas}a, it happens to propagate in the medium of Fig. \ref{plas}b. The calculation of a conformal grid for the disordered surface of Fig. \ref{plas}a requires a new formulation of optical conformal mapping, which we recently developed and allows to generate conformal grids for arbitrary structures, and with arbitrary-large numerical precision. This approach is quite involved, and will be discussed in a future work.\\
\indent The plasmonic reference of Fig. \ref{plas}b illustrates in clear form the effects of disorder, which introduces a strong modulation of the refractive index in proximity of the metallic surface at $\psi=0$, generating a network of epsilon-near-zero (ENZ) regions, separated by areas of high refractive index (Fig. \ref{plas}b). As observed in the insets of Fig. \ref{plas}a-b (dashed lines), ENZ regions are created in the points where the metallic surface is convex, while high dielectric permittivities originate in points where the surface is concave. When waves propagate into an ENZ material, their phase velocity diverges thus creating standing waves with infinite wavelength \cite{PhysRevLett.97.157403,maas13:_exper_realiz_of_epsil_near,PhysRevLett.110.013902}. When SPP waves propagate in the nanowire network of Fig. \ref{plas}a, they `see' the equivalent medium illustrated in Fig. \ref{plas}b, and get trapped in the ENZ regions thereby generating a set of quasi-localised states. We illustrate this  dynamics by a series of FDTD simulations. Figure \ref{palette}a reports a magnified version of Fig. \ref{spec}a, showing FDTD calculated reflectivity spectra for different thicknesses of the $\mathrm{Al_2O_3}$ layer. FDTD results corresponding to different combinations of alumina thicknesses and input wavelengths are summarised in Figs. \ref{palette}b-l. When light impinges on the disordered metallic structure (Fig. \ref{palette}b), some energy is scattered back generating components along all directions in space, while the remaining is coupled into SPP waves. As illustrated in Figs. \ref{palette}c-e, showing FDTD calculated electromagnetic energy density distributions, SPP waves are completely localised  in proximity of different convex points of the surface, exactly where the ENZ regions are formed. FDTD simulations show that different wavelengths are trapped in different ENZ regions of the metal, demonstrating that the ENZ network formed in Fig. \ref{plas}b does not possess a particular length scale and traps equivalently all input wavelengths. The absence of a characteristic scale is expected from the strongly disordered surface modulation of the sample, which possesses an abundant variety of different curvatures (Fig. \ref{plas}a) and therefore of ENZ regions with different extension (Fig. \ref{plas}b). The combinations of all of these ENZ regions traps polychromatic light very efficiently, as observed from the flat reflectivity response of Fig. \ref{palette}a (solid green line). In order to characterise further the energy propagation in the structure, we also plot the flow of electromagnetic energy in the structure, computed from the Poynting vector of the electromagnetic field (Fig. \ref{palette}f). This is represented with a specific line integral convolution (LIC) technique, which clearly visualises the energy flow, characterised by complex patterns with a nontrivial vorticity.\\
\indent When we deposit a small layer of $\mathrm{Al_2O_3}$ on top of the metal, the scattering dynamics changes abruptly (Fig. \ref{palette}g). In this situation, a portion of scattered wavevectors is reflected inside the alumina layer, thus generating a series of additional scattering events in the $\mathrm{Al_2O_3}$. Wavevectors associated to guided modes in the $\mathrm{Al_2O_3}$ layer are totally reflected back and do not radiate energy outside the alumina, surviving the dynamics for many scattering events.  This process creates a flow of energy in the layer of $\mathrm{Al_2O_3}$, inducing a preferential localisation of SPP inside ENZ regions that exists within the film of $\mathrm{Al_2O_3}$. Contrary to the previous case, however, ENZ regions created in the alumina have at disposal a very limited area and do not show the same variety of length scales observed in Figs. \ref{palette}c-e. This induces resonant coupling only at specific wavelengths, which resonates with the characteristic scale of the ENZ regions formed in the $\mathrm{Al_2O_3}$.  This process is clearly illustrated in Figs. \ref{palette}h-i, which show the presence of a resonant coupling around the wavelength of $425$ nm (Fig. \ref{palette}l). Resonant light localisation in ENZ regions within the alumina layer is observed in Fig. \ref{palette}i, which shows light trapping at the wavelength of $425$ nm in different points of convex metallic curvature located inside the $\mathrm{Al_2O_3}$. Outside this wavelength range, no surface localisation is formed (Fig. \ref{palette}h-i) and no reflectivity minimum is observed (Fig. \ref{palette}a, solid blue line, points h and i). Figure \ref{palette}m presents LIC images of the Poynting flux, clearly showing the flux of energy originated inside the $\mathrm{Al_2O_3}$ from the the light backscattered from the random metallic surface of the sample. The $\mathrm{Al_2O_3}$ layer, in essence, by penetrating a few nm below the metallic surface of the sample is `selecting' one particular scale in an otherwise free-scale network, thus generating an absorbing state that is macroscopically perceived with the formation of a structural colour in the material.\\  
\indent The ray optics analysis of Fig. \ref{palette}b and Fig. \ref{palette}g, beside being very intuitive to grasp, has also the advantage of providing exact results for guided modes in slab waveguides. This allows to quantitatively estimate the red-shift of the reflectivity minimum observed in both experiments and theory (Fig. \ref{spec}). This calculation is made by considering that the dispersion relation $\omega(\beta)$ of guided modes in a slab waveguide depends on the normalised frequency $V=2d\frac{\omega}{c}\sqrt{n^2-1}$, with $n$ the refractive index of alumina, $d$ the $\mathrm{Al_2O_3}$ thickness and $\beta=\frac{\omega}{c} n\sin\theta$ the propagation constant of the mode, function of the critical angle $\theta$ (Fig. \ref{palette}g). If we assume phase-matching between the guided mode and the SPP at the resonant frequency where localisation is formed, we obtain a very simple relationship for the wavelength shift $\Delta\lambda$:
\begin{equation}
\label{eq0}
\Delta\lambda=\frac{\lambda_0}{d_0}\Delta d,
\end{equation}   
where $\lambda_0$ is the wavelength of a reflectivity minimum, corresponding to a coating thickness $d_0$, and $\Delta d$ is the thickness increase of the $\mathrm{Al_2O_3}$ layer. Figure \ref{palette}n compares experimental measures with the results of Eq. (\ref{eq0}). By applying experimental values for both $\lambda_0$ and $d_0$, we obtained a coefficient $\frac{\lambda_0}{d_0}\approx 12$, which implies a wavelength shift of $12$ nm for every $1$ nm increment of coating thickness. The results of Eq. (\ref{eq0}) show a good agreement with experimental results, predicting the large red-shift that is at the basis of the structural colours formed in the system. Supplementary Fig. 6 shows FDTD calculated energy densities corresponding to different reflectivity minima formed at different thicknesses of alumina. In all cases, despite the different wavelength used for the illumination, the same set of resonant localisations is formed on the surface, validating the phase-matching condition used for the calculation of Eq. (\ref{eq0}).

\section*{Conclusions}
We have experimentally demonstrated a new design concept to create robust and highly saturated structural colours in metasurfaces composed by metallic nanowire networks with ultra-thin, lossless dielectric coatings. 
Using a combination of analytical and numerical techniques, we illustrated that these colours are the result of the resonant coupling of light with surface plasmons that are localised in equivalent epsilon-near-zero regions formed in the metallic network.
This mechanism is not constrained for large angles as high as $70^\circ$, allowing for an efficient trapping of light over a broad wavelength range in the visible. The combination of mechanical robustness and high colour saturation in an extremely light-weight structure, makes these structural colours suitable for real world industrial applications, such as automotive vehicles or airplanes for which the weight is directly related to the fuel economy. As discussed in the introduction, achieving a scalable fabrication is a key problem in structural colour printing. On the basis of our experiments, it is evident that our metasurfaces have shown a wide colour capability without the need of electron beam lithography (EBL) or other complex fabrication procedures. Our structures, in fact, are based on simple wet-chemistry and coating technologies, which can produce robust colours on very large spatial scales.\\
In addition to such fundamental advances, our design concept has the potential to enrich the application of metasurfaces to areas where large active regions are mandatory, such as efficient light trapping layers in photovoltaics cells. Although a deeper discussion on this topic goes beyond the scope of this paper, we can here introduce some important points. On the basis of our theory and experiments, we demonstrated that is possible to control the  response of an optical material by `engineering' the connectivity of a network of ENZ nanostructures created in a random metallic structure. From the results of Fig. \ref{palette}, we observe that this approach allows to strongly localise optical radiation in nanoscale regions located well outside the metal, completely absorbing incoming optical photons in a specific bandwidth (Fig. \ref{palette}l-m). This approach can potentially enhance the absorption power of ultra-thin absorbers, which can take advantage from the formation of localised spots and harvest a significant portion of light energy in nm-thick film structures. The current photovoltaic technology employs Si absorbers of approximatively $100$~$\mu$m thickness, while other solution processed materials with high manufacturability and low cost, such as quantum dots, require films thickness larger than $1$~$\mu$m to efficiently absorbs all incoming photons. Our metastructures can considerably scale down these thicknesses, stimulating new research aimed at developing innovative materials for renewable energy harvesting. 

\subsection*{Acknowledgments}
For the computer time, we used the resources of the KAUST Supercomputing Laboratory and the Redragon cluster of the Primalight group.\\ 
This work was performed in part at the Center for Nanoscale Systems (CNS), a member of the National Nanotechnology Infrastructure Network (NNIN), which is supported by the National Science Foundation under NSF award no. ECS-0335765. CNS is part of Harvard University.\\
H. Galinski acknowledges the financial support of the `Size matters'-project (TDA Capital Ltd, London, UK). H. Dong acknowledges the financial support by the Master Thesis Grant of the Zeno Karl Schindler Foundation (Switzerland).

\newpage
\section*{Figure Legends}

Fig. 1. \textbf{Observation of structural colors in random metallic networks with subwavelength dielectric coatings}. (\textbf{a}) Schematic illustration of an Al$_2$O$_3$ coated PtYAl nanomaterial, based on a three dimensional (3D) reconstruction of a completely dealloyed Pt-Y-Al thin-film obtained via focused ion beam (FIB) assisted thin-film tomography.  (\textbf{b}) Photographs of as deposited, dealloyed and Al$_2$O$_3$ coated PtYAl metamaterial-networks, illustrating the formation of vibrant colors and the continuous color change with increasing coating thickness. The photographs have been taken under illumination from ceiling lights. Each image is $2\times2$~mm$^2$. (\textbf{c}) Experimental and FDTD simulated structural color reported in a standard CIE 1931 $(x,y)$ space, depicting all of the chromaticities visible to the average person. The RGB color space is marked by the triangle area. The chromaticity is calculated directly from reflectance spectra obtained either experimentally (circles markers) or by FDTD simulations (dashed line). The edges of the tongue-shaped plane correspond to color values of maximal saturation. \textbf{d}, Color palette of generated colors calculated by FDTD simulations for increasing thickness of Al$_2$O$_3$.\\

Fig. 2. \textbf{Examples of different graphic arts design with structural colors from metamaterial networks}. Photograph and optical micrograph of a colored graphic art designed by combining a RF-sputtered Al$_2$O$_3$ coated network metamaterial and photolithography. The inset shows an optical micrograph illustrating a detail of the graphic art and the uniformity of the color.\\

Fig. 3. \textbf{Wear properties of the structural colors}. (\textbf{a}) Schematic illustration of the  basic principle of the scratch testing technique. A diamond stylus is used to scratch the film with progressively increasing load. (\textbf{b}) Optical micrographs of progressive load scratches ($0.4-15$mN) on a dense PtYAl thin film with and without coating and PtYAl nanoscale networks coated with $28$ and $53$~nm thick Al$_2$O$_3$, respectively. The critical load characterizing the adhesion failure of the films is indicated by a pink arrow.\\

Fig. 4. \textbf{Optical properties of the network metamaterials: reflectivity spectra.} (\textbf{a}) Experimental and FDTD calculated normal incidence reflectance spectra as function of the Al$_2$O$_3$ coating thickness $d$. 
 (\textbf{b-e}) Experimental reflectance spectra of nanoporous PtYAl thin films coated with 18~nm, 28~nm, 45~nm and 53~nm Al$_2$O$_3$, respectively, as function of the incidence angle ($20^{\circ}$-$85^{\circ}$). The value of reflectance is indicated by the color bar.\\

\begin{figure*}[t]
      \begin{center}
		\includegraphics[width=0.90\textwidth]{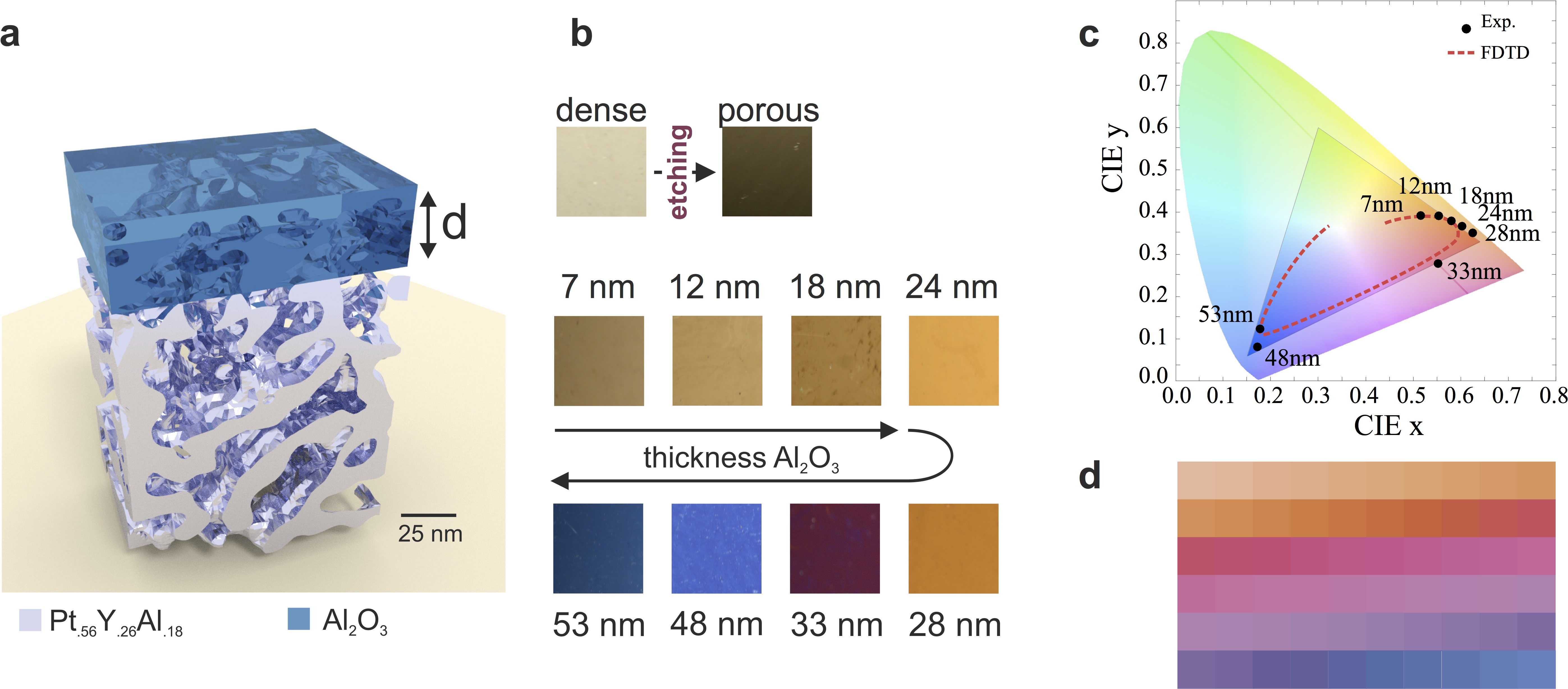}
		\end{center}
\caption{
}
  \label{pal}
\end{figure*}

\begin{figure*}
\begin{center}
		\includegraphics[width=9cm]{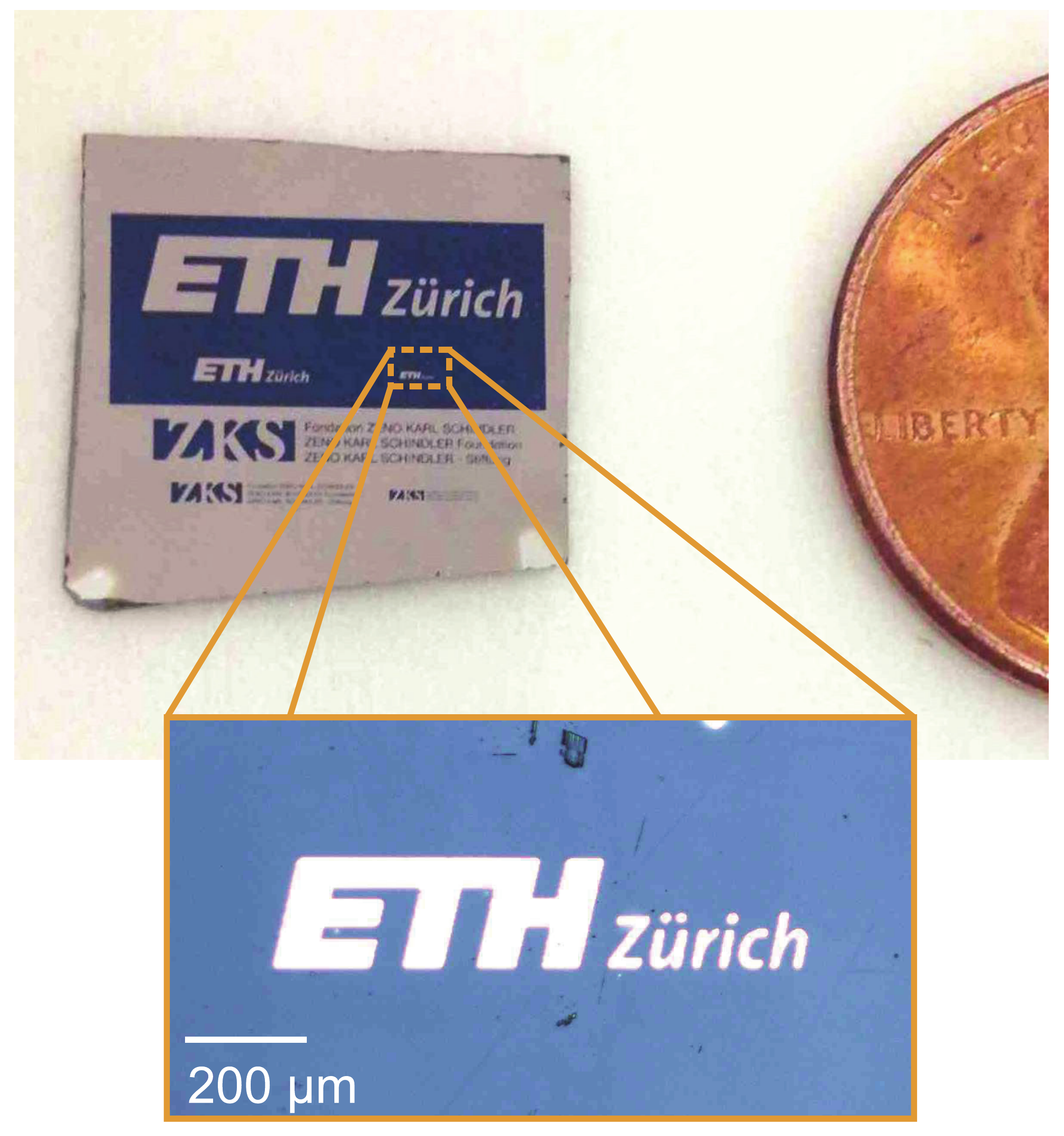}
\end{center}
\caption{
}
\label{art}
\end{figure*}

\begin{figure*}
\begin{center}
		\includegraphics[width=0.95\textwidth]{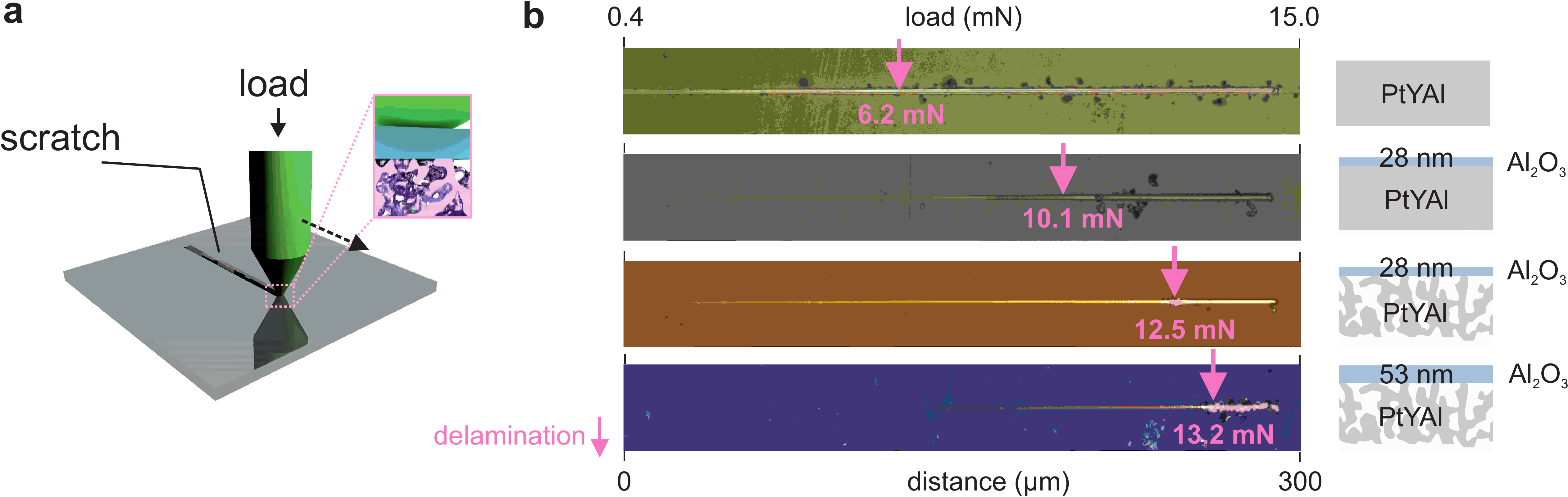}
\end{center}
\caption{
}
\label{wear}
\end{figure*}

Fig. 5. \textbf{Generation of an equivalent epsilon-near-zero material in the metallic nanowire network of Fig. \ref{pal}a.} (\textbf{a}) 2D cross section profile of the metallic nanowire network, as obtained from experimental FIB images of polished samples (Fig. \ref{pal}a). When light impinges on this structure, it excites the propagation of surface plasmon polariton waves (SPP), which move along the complex surface of the metal (Panel a, inset). This motion is conveniently described in a curvilinear reference $(\phi,\psi)$, which provides a conformal map of the metallic surface of the sample (solid red line). In the transformed space $(\phi,\psi)$ (panel b), SPP waves appear to propagate inside an inhomogenous material with refractive index $n(\phi,\psi)$, on the line at $\psi=0$ (c, inset). The material $n(\phi,\psi)$ models the effects of the metallic geometry of panel (a), which is flattened out in transformed space $(\phi,\psi)$. The two systems of panels a-b are exactly the same for light propagation. The equivalent structure of (b) shows a complex network of ENZ structures (panel b, dark blue area), which are created by points of convex metallic curvature (right inset).\\

\begin{figure*}
      \begin{center}
		\includegraphics[width=0.99\textwidth]{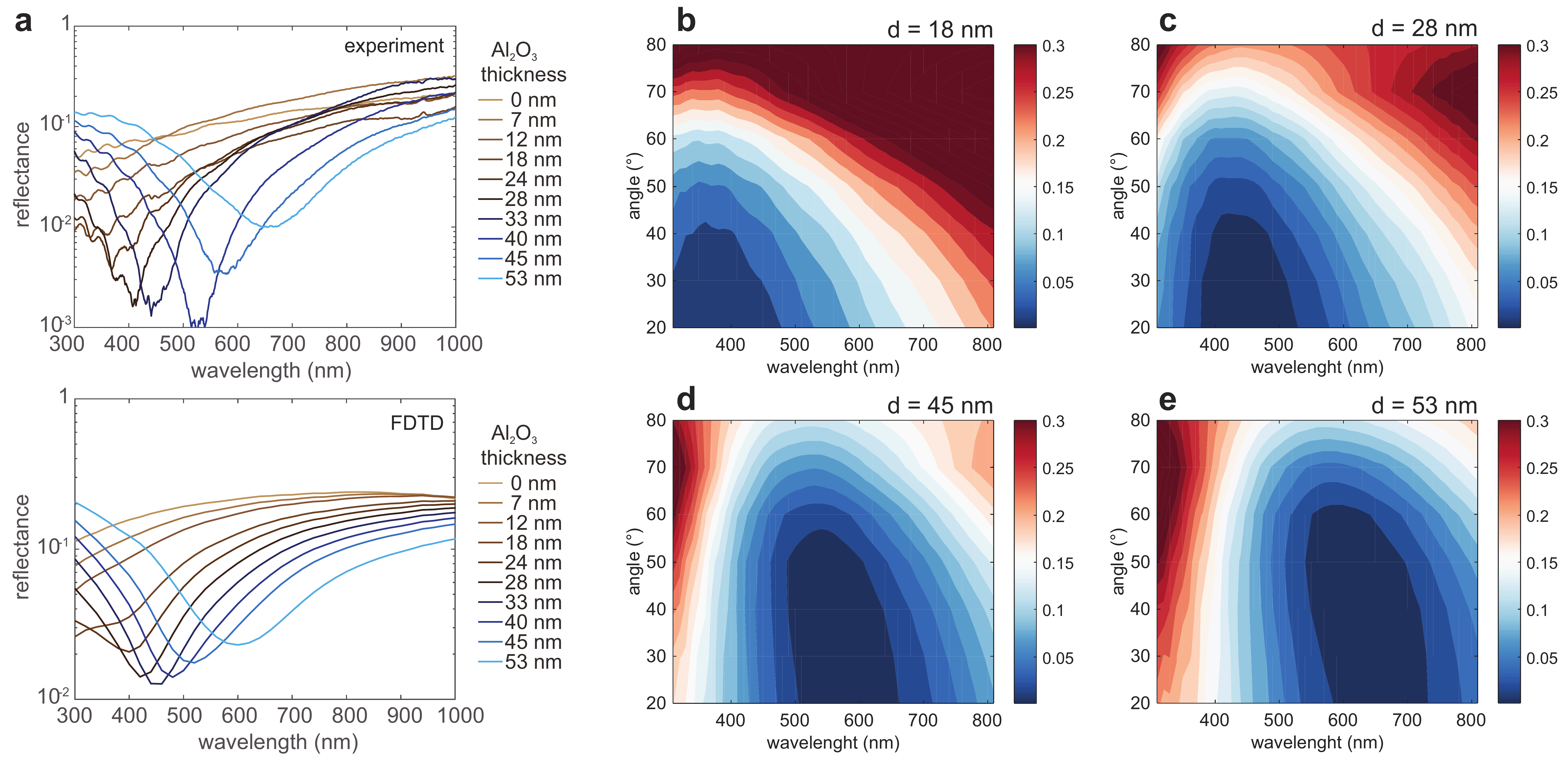}
		\end{center}
 \caption{
 }
  \label{spec}
\end{figure*}

\begin{figure*}
      \begin{center}
		\includegraphics[width=0.99\textwidth]{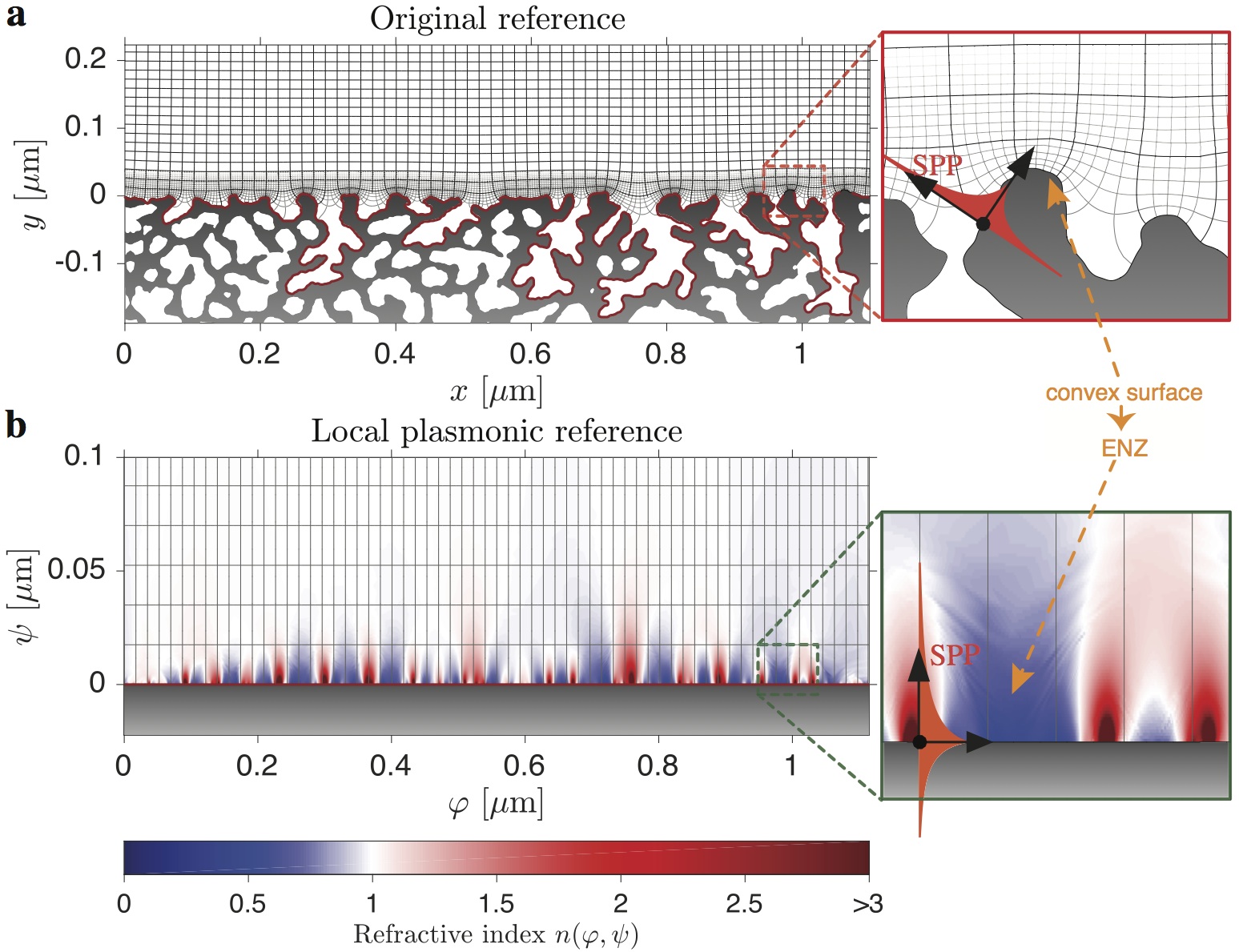}
		\end{center}
  \caption{
  }
  \label{plas}
\end{figure*}

\begin{figure*}[t!]
  \centering
  \includegraphics[width=0.99\textwidth]{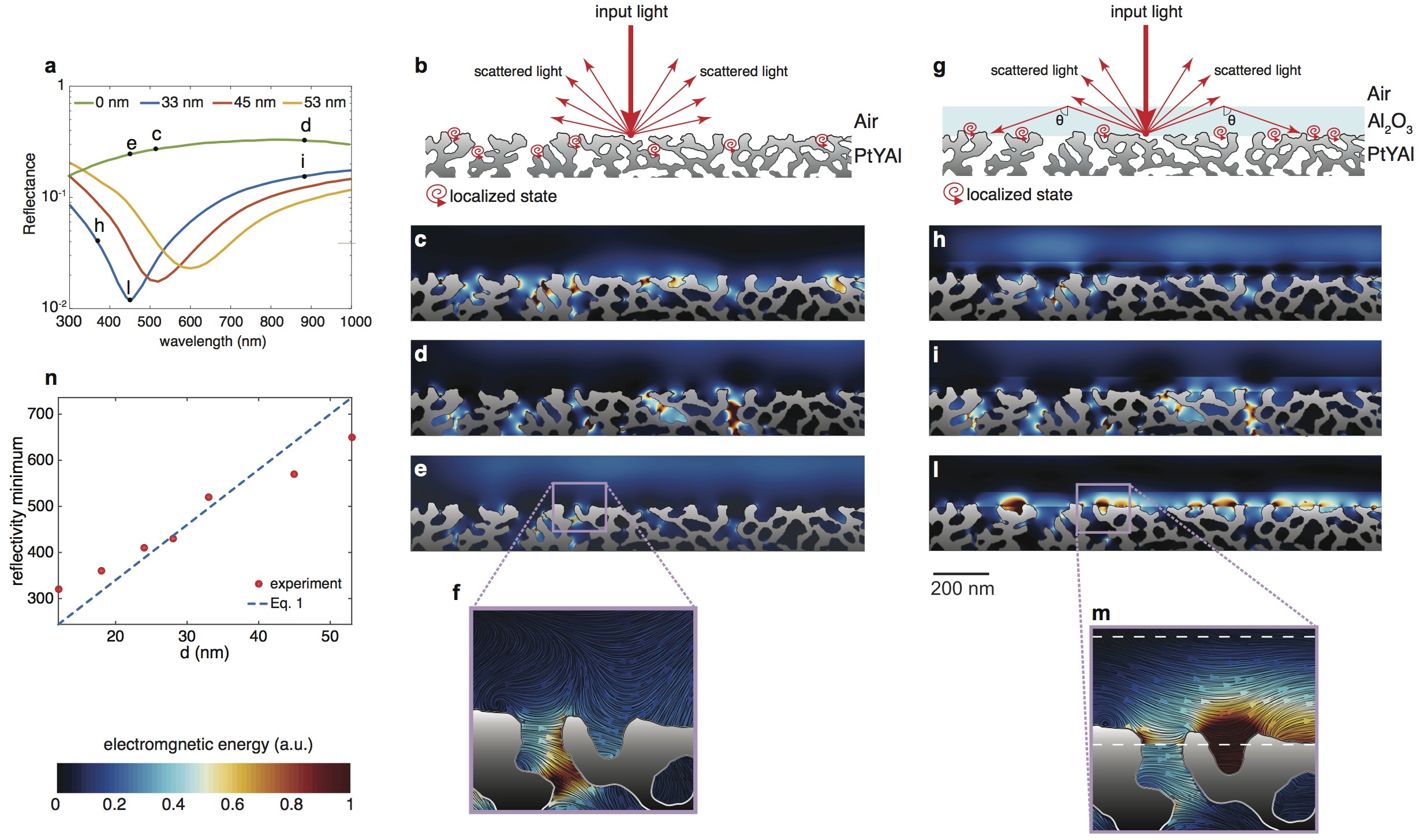}
  \caption{
  }
\label{palette}
\end{figure*}

Fig. 6. \textbf{Mechanisms of structural colours formation in the PtYAl cellular network.} (a) Normal-incidence reflectance spectra obtained from FDTD simulations of the nanoscale Pt-network of Fig. \ref{pal}a with different thickness of Al$_2$O$_3$. Panels (b-f) analyse the case with no Al$_2$O$_3$ deposited on top of the metal, while panels g-m summarise the results for an Al$_2$O$_3$ layer of $33$ nm. Panels (b,g) provide a pictorial illustration of light-matter interaction with the sample, without (b) and with (g) Al$_2$O$_3$. In the presence of Al$_2$O$_3$, a portion of scattered waves is reflected back in the Al$_2$O$_3$ layer, thus creating an energy flow in the coating layer and a resonant coupling with ENZ regions located in the Al$_2$O$_3$. Panels (c-e) and (h-l) present FDTD calculated spatial energy distributions in the structure by considering an input wavelength indicated by the corresponding letter in panel a. Energy distributions are averaged over one optical cycle at steady state. Panel (f,m), conversely, provide a zoomed view of the pink area of panel (e,l), and illustrate the electromagnetic energy flow in the structure (arrow coloured lines). The flow is superimposed with the corresponding averaged spatial energy distribution. Panel n, finally, compares the reflectivity minimum shift observed in experiments (Fig. \ref{spec}) with theoretical predictions based on the model illustrated in panel g.

\bibliographystyle{vancouver}
\bibliography{refbibNew}



\newpage

\clearpage

\clearpage

\clearpage

\end{document}